\documentclass[reprint,twocolumn,nofootinbib,superscriptaddress]{revtex4}

\usepackage{graphicx}%
\usepackage{dcolumn}%
\usepackage{bm}%
\usepackage{epsfig,colordvi}
\usepackage[section]{placeins}

\usepackage{xcolor}

\begin{document}


\title{Constraints on the kinetic mixing parameter $\epsilon^2$ for the light dark photons
  from dilepton production in heavy-ion collisions in the few-GeV energy range}


\author{Ida Schmidt}
\affiliation{Institut f\"ur Theoretische Physik, Johann Wolfgang Goethe-Universit\"at,
Max-von-Laue-Str. 1, 60438 Frankfurt am Main, Germany}

\author{Elena~Bratkovskaya}\email{E.Bratkovskaya@gsi.de}
 \affiliation{GSI Helmholtzzentrum f\"ur Schwerionenforschung GmbH,
  Planckstr. 1, 64291 Darmstadt, Germany}
 \affiliation{Institut f\"ur Theoretische Physik, Johann Wolfgang Goethe-Universit\"at,
Max-von-Laue-Str. 1, 60438 Frankfurt am Main, Germany}
\affiliation{Helmholtz Research Academy Hessen for FAIR (HFHF), GSI Helmholtz Center for Heavy Ion Physics, Campus Frankfurt, 60438 Frankfurt, Germany}

\author{ Malgorzata Gumberidze}
\affiliation{GSI Helmholtzzentrum f\"ur Schwerionenforschung GmbH,
  Planckstr. 1, 64291 Darmstadt, Germany}

\author{ Romain Holzmann}
\affiliation{GSI Helmholtzzentrum f\"ur Schwerionenforschung GmbH,
  Planckstr. 1, 64291 Darmstadt, Germany}


\begin{abstract}
The vector $U$-bosons, or so called 'dark photons', are one of the possible candidates for the dark matter mediators.
They are supposed to interact with the standard matter via a 'vector portal' due to the $U(1)-U(1)^\prime$ symmetry group mixing  
which might make them visible in particle and heavy-ion experiments.
While there is no confirmed observation of dark photons, the detailed analysis of different experimental data allows to 
estimate the upper limit for the kinetic mixing parameter $\epsilon^2$  depending on the mass $M_U$ of $U$-bosons 
which is also unknown. 
In this study we present theoretical constraints on the upper limit of $\epsilon^2(M_U)$ in the mass range 
$M_U \le 0.6$ GeV from the comparison of the calculated dilepton spectra with the experimental data from 
the HADES Collaboration at SIS18 energies where the dark photons are not observed.  
Our analysis is based on the microscopic Parton-Hadron-String Dynamics (PHSD) transport approach
which reproduces well the measured dilepton spectra in $p+p$, $p+A$ and $A+A$ collisions. Additionally to the
different dilepton channels originating from interactions and decays of ordinary matter particles (mesons and baryons), 
we incorporate the decay  of hypothetical $U$-bosons to dileptons, $U\to e^+e^-$, where the $U$-bosons themselves
are produced by the Dalitz decay of pions $\pi^0\to \gamma U$, $\eta$-mesons
$\eta \to \gamma U$ and Delta resonances $\Delta \to N U$. Our analysis can help to estimate the requested 
accuracy for future experimental searches of 'light' dark photons by dilepton experiments.
\end{abstract}


\maketitle


\section{Introduction}

An understanding of the structure of our Universe is one of the intriguing topics of modern physics.
According to the present knowledge, the standard matter represents less then 5\% of our Universe, while
about 27\% of it consists of  so-called 'dark matter' (DM) and about 68\% is the 'dark energy' \cite{NASADM21}.
The dark matter is supposed to be a relic from the Big Bang, which makes itself noticeable by its gravitational
action on the large-scale cosmic structures.
It was advocated that the dark matter mediators can interact with the Standard Model (SM) particles 
by four possible 'portals'  -- vector, Higgs, neutrino and axion (cf. the reviews 
\cite{Alexander:2016aln,Battaglieri:2017aum,Agrawal:2021dbo} and references therein).  

The 'vector' portal assumes the existence of a  $U(1)-U(1)^\prime$ gauge symmetry 
group mixing \cite{Holdom:1985ag}, i.e. the corresponding Lagrangian is
defined by the hypercharge field-strength tensor of the SM photon field and the DM vector boson field:
${\cal L} \sim \epsilon^2/2 \, F_{\mu\nu}{F^{\mu\nu}}^\prime$. 
The mediators in this case are vector $U$-bosons, which often are called 'dark' or 'hidden' photons or $A^\prime$,
with a mass $M_U$ remaining presently unknown. 
Here $\epsilon^2$ is a kinetic mixing parameter, which characterizes the strength of the interaction of SM and DM
particles \cite{Fayet:1980ad,Fayet:2004bw,Boehm:2003hm,Pospelov:2007mp,Batell:2009di,Batell:2009yf}.
 This mixing allows the decay of $U$-bosons to a pair of leptons - $e^+e^-$ or $\mu^+\mu^-$.
The 'light' $U$-bosons can be produced by the decay of SM particles, e.g. by Dalitz decays of pseudoscalar mesons - pions 
and $\eta$-mesons, as well as by the Dalitz decay of baryonic resonances such as $\Delta$'s.
This provides the possibility to observe  dark photons in dilepton experiments, stimulating a lot of experimental 
as well as theoretical activities \cite{Alexander:2016aln,Battaglieri:2017aum,Beacham:2019nyx,Billard:2021uyg}.
A recent measurement of the excess electronic recoil events by the XENON1T Collaboration might be interpreted also in 
favour of dark matter sources and in particular dark photons are possible candidates \cite{Aprile:2020tmw}. 

The HADES Collaboration at GSI, Darmstadt, performed an experimental search for dark photons in dilepton experiments
at SIS18\footnote{The SIS18 accelerator at GSI, Darmstadt delivers protons with kinetic energies up to 4.5 GeV and
heavy-ion beams with kinetic energies per nucleon up to 2 GeV  (denoted as A GeV).} with both proton and heavy-ion beams \cite{Agakishiev:2013fwl}.
The HADES experiment presented an upper limit for the kinetic mixing parameter $\epsilon^2$ in the mass range of
$M_U=0.02-0.55$ GeV based on the experimental measurements of $e^+e^-$ pairs from $p+p$ and $p+Nb$ collisions
at 3.5 GeV as well as $Ar+KCl$ collisions at 1.76 $A$GeV.  
Later the HADES result has been superseded by the A1 \cite{Merkel:2014avp}, the NA48/2 \cite{Batley:2015lha} and the BaBar \cite{Lees:2014xha,Lees:2017lec} experiments which further lowered the limit for $\epsilon^2$ in this mass range.
The NA48/2 experiment investigated a large sample of $\pi^0$ Dalitz decays obtained from in-flight weak kaon decays, the BaBar collider experiment used their cumulated luminosity of $e^+e^-$  reactions to survey a very large mass range up to $M_U = 8$ GeV, and the A1 experiment \cite{Merkel:2014avp} at MAMI investigated electron scattering off a $^{181}Ta$ target
at energies between 180 and 855 MeV to search for a dark photon signal.  
In the mass range discussed here, i.e. 20 - 500 MeV, the limit on $\epsilon^2$ has thus  been pushed down to about $10^{-6}$.  
The compilation of the world data collected by various experiments can be found in the review \cite{Battaglieri:2017aum}.

The goal of this study is to estimate the upper limit for the kinetic mixing parameter $\epsilon^2(M_U)$ 
depending on the mass of the hypothetical $U$-boson from the theoretically calculated
dilepton spectra using the microscopic Parton-Hadron-String Dynamics (PHSD) transport approach which
describes the whole evolution of heavy-ion collisions based on microscopic transport theory by solving the equations-of-motion
for each degree-of-freedom (hadronic and partonic) and their interactions.  The PHSD provides
a consistent description of hadron production in $p+p$, $p+A$ and $A+A$ collisions 
as well as electromagnetic probes - dileptons and photons, from SIS18 to LHC energies 
(cf.  \cite{Linnyk:2015rco,Linnyk:2012pu,Song:2018xca}).
In particular, the PHSD describes very well the dilepton data of the HADES experiment \cite{Bratkovskaya:2013vx}.
Having the SM particle production under control, we incorporated in the PHSD the dynamical production 
of $U$-bosons by $\pi^0, \eta$ and $\Delta$ Dalitz decays as well as the $U$-boson decay to $e^+e^-$ pairs. 
We compare our results with the experimental data from the HADES Collaboration and provide constraints 
on the mass dependence of the kinetic mixing parameter $\epsilon^2$ for light dark photons ($M_U \le 0.6$ GeV).

Our paper is organised as follows:
In Section II we recall the basic ideas of the PHSD approach and the SM production. In Section III
we describe the production of dark photons and in Section IV we show the results for the mixing parameter $\epsilon^2(M_U)$.
We summarise our findings in Section V.

\section{Standard matter (SM) production in the PHSD}

In this section we recall the basic ideas of the PHSD approach and treatment of the dilepton production 
from  SM matter. 

\subsection{The PHSD transport approach }

The Parton-Hadron-String Dynamics is a non-equilibrium microscopic transport approach  that incorporates hadronic as well as partonic degrees-of-freedom 
\cite{Cassing:2008sv,Cassing:2008nn,Cassing:2009vt,Bratkovskaya:2011wp,Konchakovski:2011qa,Linnyk:2015rco,Moreau:2019vhw}. 
The PHSD describes the full evolution of a relativistic heavy-ion collision, from the initial hard NN collisions  
out-of equilibrium, to the formation of the quark-gluon plasma (QGP), its partonic interactions, 
up to the hadronisation and final-state interactions of the resulting hadrons.
The dynamical description of the time evolution of the interacting system is based on the solution of the Cassing-Juchem generalised off-shell transport equations in test-particle representation  \cite{Cassing:1999wx,Cassing:1999mh}  on the basis of the  Kadanoff-Baym equations \cite{KadanoffBaym} in first-order gradient expansion, which are applicable for the dynamical description of strongly interacting degrees-of-freedom \cite{Juchem:2004cs,Cassing:2008nn}.

The hadronic sector follows the early developments from the HSD transport approach \cite{Ehehalt:1996uq,Cassing:1999es} which includes as explicit hadronic degrees-of-freedom the baryon octet and decouplet, the ${0}^{-}$ and ${1}^{-}$ meson nonets and higher resonances.  The description of multi-particle production in elementary baryon-baryon ($BB$),
meson-baryon ($mB$) and meson-meson ($mm$) reactions is incorporated based on the Lund model \cite{NilssonAlmqvist:1986rx}.
This is realized  in terms of the  event generators FRITIOF 7.02 \cite{NilssonAlmqvist:1986rx,Andersson:1992iq} and PYTHIA 6.4 \cite{Sjostrand:2006za} which are  "tuned" (cf. Ref. \cite{Kireyeu:2020wou} for details)
for a better description of elementary reactions at low and intermediate energies
as well as for the incorporation of the  in-medium effects related to chiral symmetry restoration and modification of the properties of the formed hadronic degrees-of-freedom, while the string formation and decay occurs in a dense hadronic medium \cite{Cassing:2015owa,Palmese:2016rtq,Bratkovskaya:2007jk,Cassing:2003vz,Song:2020clw}.
The strings are built from primary nucleon-nucleon ($NN$) collisions and secondary energetic $BB$, $mB$ and $mm$ interactions
(the corresponding thresholds are taken as $(s_{BB}^{th})^{1/2}=2.65$ GeV,  $(s_{mB}^{th})^{1/2}=2.4$ GeV and  
$(s_{mm}^{th})^{1/2}=1.3$ GeV).
They decay to the leading hadrons (the energetic ends of the strings) and 'pre-hadrons', i.e. the newly produced 
mesons and baryons which are considered under formation time $t_F=\tau_F\gamma$ (where $\gamma=1/\sqrt{1-v^2}$
and  $v$ is the velocity of the particle in the calculational frame which is chosen to be the initial 
$NN$ center-of-mass frame).

The transition from hadronic to partonic degrees-of-freedon and vice versa (i.e. hadronization) occurs when
the local energy density is above (below) the critical energy density of $\epsilon_C \sim 0.5$~GeV/${fm}^{3}$ 
in line with lattice quantum chromodynamics (lQCD) \cite{Borsanyi:2015waa}. If the energy density is below critical the  'pre-hadrons' evolve into
asymptotic hadronic states after the formation time $t_F$ and interact with hadronic cross sections. 

The description of the partonic degrees-of-freedom and their interactions during the QGP phase is based on the Dynamical Quasi-Particle Model (DQPM) \cite{Cassing:2007yg,Cassing:2007nb} which describes the thermodynamic properties of QCD in equilibrium in terms of massive strongly-interacting quasi-particles whose masses are distributed according to spectral functions (imaginary parts of the complex propagators).  The widths and pole masses of the spectral functions are defined by the real and imaginary parts of the parton  self-energies and the effective coupling strength in the DQPM; both depend on the local temperature $T$ (or local energy density) and the baryon chemical potential $\mu_B$. They are fixed 
by adjusting the DQPM entropy density to the respective lQCD results from Refs. \cite{Aoki:2009sc,Cheng:2007jq}.
The QGP phase is evolved by the off-shell transport equations with self-energies and cross sections from the DQPM. When the fireball expands the probability of the partons for hadronization increases close to the phase boundary 
between hadronic and partonic phases (which is a crossover at all RHIC energies); the hadronisation takes place using covariant transition rates. The resulting hadronic system is further-on governed by the off-shell HSD dynamics incorporating (optionally) self-energies for the hadronic degrees-of-freedom \cite{Cassing:2003vz}.

We recall that the PHSD approach has been successfully employed for $p+p$, $p+A$ and $A+A$ collisions from 
SIS18 to LHC energies and reproduces many hadronic observables as well as the dilepton and photon observables 
(cf. Refs. \cite{Cassing:2008sv,Cassing:2008nn,Cassing:2009vt,Bratkovskaya:2011wp,Konchakovski:2011qa,Linnyk:2015rco,Moreau:2019vhw,Song:2018xca}). 

\subsection{Dilepton production from the SM sources in the PHSD}

Dileptons ($e^+e^-$, $\mu^+\mu^-$ pairs) produced from the decay of  virtual photons can be
emitted from all stages of the heavy-ion reactions from hadronic and partonic sources 
\cite{Rapp:2013nxa,Linnyk:2015rco}.
The hadronic sources at low invariant masses ($M < 1$ GeV$c$) are the  Dalitz decays
of mesons and  baryons $(\pi^0,\eta,\Delta, ...)$ and the direct decays of
vector mesons  $(\rho, \omega, \phi)$ as well as hadronic bremsstrahlung; 
at intermediate masses (1 GeV$ < M < 3$ GeV) the leptons issue from semileptonic decays of 
correlated $D+\bar D$ pairs as well as from 
multi-meson reactions ($\pi+\pi, \ \pi+\rho, \ \pi+\omega, \
\rho+\rho, \ \pi+a_1, ... $) denoted by ``$4\pi$" contributions; 
at high invariant masses ($M > 3$ GeV) the main sources are the direct decay of
vector mesons  $(J/\Psi, \Psi^\prime)$ and
initial 'hard' Drell-Yan annihilation to dileptons ($q+\bar q \to l^+ +l^-$, where $l=e,\mu$).
The partonic sources are related to the 'thermal' QGP dileptons radiated from the partonic interactions
where the leading processes are the 'thermal' $q\bar
q$ annihilation ($q+\bar q \to l^+ +l^-$, \ \ $q+\bar q \to g+ l^+ +l^-$) 
and Compton scattering ($q(\bar q) + g \to q(\bar q) + l^+ +l^-$) in the QGP; 
they contribute dominantly to the intermediate mass regime.
 
Since in this study we concentrate on low energy heavy-ion collisions, only hadronic sources are relevant for the considerations here.
For the description  of dilepton production from the QGP sources we refer the reader to the review
\cite{Linnyk:2015rco} and to the recent study on the relative contribution
of the QGP dileptons and those from the correlated charm decays \cite{Song:2018xca}.

The dilepton production by a (baryonic or mesonic) resonance $R$
decay can be schematically presented in the following way:
\begin{eqnarray}
 BB &\to&R X   \label{chBBR} \\
 mB &\to&R X \label{chmBR} \\
 mm &\to&R X \label{chmmR} \\
       R & \to & e^+e^-, \label{chR} \\
       R & \to & e^+e^- X, \label{chRd} \\
       R & \to & m X, \ m\to e^+e^- X, \label{chRMd} \\
       R & \to & R^\prime X, \ R^\prime \to e^+e^- X, \label{chRprd}
\end{eqnarray}
i.e. in a first step a resonance $R$ might be produced in
baryon-baryon (\ref{chBBR}), meson-baryon  (\ref{chmBR}) or meson-meson  (\ref{chmmR})
collisions (or at high energy collisions the resonance can be formed in the hadronization process). 
Then this resonance can decay to dileptons directly, e.g.  a direct decay  
of vector mesons ($\rho, \omega, \phi$). 
The decay of resonances to dileptons can be accompanied by the production of other particles (\ref{chRd}) 
(e.g., Dalitz decay of the $\Delta$ resonance: $\Delta
\to e^+e^-N$). It can decay firstly to a meson $m$ (+ hadrons) and later to dileptons (\ref{chRMd})  
by e.g. the Dalitz decay ($\pi^0, \eta, \omega$). The resonance $R$ might
also decay into another resonance $R^\prime$ (\ref{chRprd}) which
later produces dileptons via Dalitz decay.

The electromagnetic part of all conventional dilepton sources  --
$\pi^0, \eta, \omega, \Delta$  Dalitz decays as well as direct decay of vector
mesons $\rho, \omega$ and $\phi$ -- are calculated following our early work \cite{Bratkovskaya:2000mb}. 
However, here we adopt the "Wolf model" for the differential electromagnetic decay width of the $\Delta$ resonance
\cite{Wolf:1990ur} which is a default setting in the PHSD 4.0 used for this study.
The details of the evaluation of the $\Delta$ Dalitz decays are given in Ref. 
\cite{Bratkovskaya:2013vx}.

For the bremsstrahlung from $pp$ and $pn$ reactions ($NN\to NN\gamma^* \to NN e^+e^-$)
as well as from $\pi N$ 'quasi-elastic' scattering ($\pi N\to \pi N\gamma^* \to \pi N e^+e^-$)
we adopt the results from the OBE model calculations by Kaptari and K\"ampfer in Ref.
\cite{Kaptari:2005qz} as implemented  in the PHSD in Ref. \cite{Bratkovskaya:2007jk}
and used for the dilepton study at SIS18 energies in Ref. \cite{Bratkovskaya:2013vx}.

We note that we account for the in-medium effects for the vector meson dynamics  
such as a 'collisional broadening' scenario for the spectral functions  \cite{Bratkovskaya:2007jk}.
The vector meson production and propagation follow 
the off-shell transport description \cite{Juchem:2004cs,Cassing:2007nb} where
their spectral functions change dynamically during the propagation through
the medium and evolve towards on-shell spectral function{s} in the vacuum.  
The effect of collisional broadening of the vector-meson spectral functions is incorporated 
by a modification of the vector meson width in the dense baryonic medium:
\begin{eqnarray}
\Gamma^*_V(M,|\vec p|,\rho_N)=\Gamma_V(M) + \Gamma_{coll}(M,|\vec
p|,\rho_N) . \label{gammas}
\end{eqnarray}
Here $\Gamma_V(M)$ is the total width of the vector mesons
($V=\rho, \omega, \phi$) in the vacuum.
The collisional width in (\ref{gammas}) is approximated as
\begin{eqnarray}
\Gamma_{coll}(M,|\vec p|,\rho_N) = \gamma \ \rho_N \langle v \
\sigma_{VN}^{tot} \rangle \approx  \ \alpha_{coll} \ \frac{\rho_N}{\rho_0}
. \label{dgamma}
\end{eqnarray}
Here $v=|{\vec p}|/E; \ {\vec p}, \ E$ are the velocity, 3-momentum
and energy of the vector meson in the rest frame of the nucleon
current and $\gamma^2=1/(1-v^2)$; $\rho_N$ is the
nuclear density and $\sigma_{VN}^{tot}$ the meson-nucleon total
cross section. We use the 'broadening coefficients'
$\alpha_{coll} \approx 150$~MeV for the $\rho$ and $\alpha_{coll} \approx 70$~MeV
for $\omega$ mesons as obtained in Ref. \cite{Bratkovskaya:2007jk}.
We note that the collisional broadening scenario is supported by a comparison of transport 
model calculations with dilepton data from low energies 
\cite{Bratkovskaya:2007jk,Bratkovskaya:2013vx,Galatyuk:2015pkq,Staudenmaier:2017vtq,Larionov:2020fnu}
to ultra-relativistic energies (cf. e.g. \cite{Santini:2011zw,Linnyk:2011vx,Endres:2016tkg,Linnyk:2015rco}).

The dilepton yield is calculated using the time integration (or 'shining') method, i.e. 
the virtual emission of dileptons by different sources is accommodated in the dilepton rates at each time
step.  In this way, the vector mesons and baryon resonances continously 'emit' dileptons from their
production ('birth') up to their absorption or hadronic decay ('death'). 
This is especially important for the study of in-medium effects because this
method takes the full in-medium dynamics into account.

\section{U-boson production and their dilepton decay in the PHSD}

Now we step to the description of $e^+e^-$ pair production from $U$-boson decays $U\to e^+e^-$ 
in the PHSD approach. 
We note that at SIS18 energies, which are in the focus of our present study, the dominant production channels of 
dileptons for $M <0.6$ GeV are the Dalitz decays of $\pi^0$, $\eta$, and the $\Delta$ resonance 
\cite{Bratkovskaya:2007jk,Bratkovskaya:2013vx}.
Thus, one expects that if the hypothetical dark photons have a mass $M_U < 0.6$ GeV, they might stem from the
Dalitz decays of $\pi^0$, $\eta$, and the $\Delta$ resonance, too. The evaluation of the corresponding
partial decay widths of pseudoscalar mesons and $\Delta$ baryons to $U$-bosons due to the $U(1)-U(1)^\prime$ mixing
has been performed in Refs. \cite{Batell:2009yf,Batell:2009di} and employed also in the HADES 
experimental search for dark photons in Ref. \cite{Agakishiev:2013fwl}.

In this study we follow the strategy  of  Ref. \cite{Agakishiev:2013fwl} and
consider three production channels of dark photons which are dominant for low energy heavy-ion 
collisions as measured by the HADES Collaboration  \cite{Agakishiev:2013fwl}, i.e.  
dark photons from Dalitz decay of 
1) neutral pions:  $\pi^0 \to \gamma + U$, 2) and $\eta$-mesons:  $\eta \to \gamma + U$,  
3) $\Delta$ resonances:  $\Delta \to N + U$, where  $U\to e^+e^-$.

The dilepton yield from a $U$-boson decay of mass $M_U$ can be evaluated as the sum of all possible contributions 
(for a given mass $M_U$):
\begin{eqnarray}
N^{U\to e^+e^-} = N^{U\to e^+e^-}_{\pi^0}+N^{U\to e^+e^-}_{\eta}+N^{U\to e^+e^-}_{\Delta} \label{NUee} \\
 = Br^{U\to e^+e^-} (N_{\pi^0\to \gamma U}+N_{\eta\to \gamma U}
+N_{\Delta\to N U}), \nonumber
\end{eqnarray}
where $Br^{U\to e^+e^-}$ is the branching ratio for the decay of $U$-bosons to $e^+e^-$. 
In this study we assume that the width of the $U$-boson is zero (or very small), i.e. it contributes only 
to a single $dM$ bin of dilepton spectra from SM sources. Accordingly, if $M_U > m_\eta$, only the $\Delta$
channel is kinematically possible in Eq. (\ref{NUee}). 

On the other hand,  the yield of $U$-bosons  of mass $M_U$ themselves can be
estimated from the coupling to $\pi^0$, $\eta$ and $\Delta$ decays to the virtual photons \cite{Agakishiev:2013fwl}:
\begin{eqnarray}
&& N_{m\to \gamma U} = N_m Br_{m\to \gamma \gamma}  \cdot
   \frac{\Gamma_{m\rightarrow\gamma U}}{\Gamma_{m\rightarrow\gamma\gamma}}, \ \ m=\pi^0, \eta  \label{mNU1} \\
&& N_{\Delta\to N U} = N_\Delta Br_{\Delta\to N \gamma}  \cdot    
   \frac{\Gamma_{\Delta\rightarrow NU}}{\Gamma_{\Delta\rightarrow N\gamma}}. \label{mNU2}
\end{eqnarray}

Following Refs. \cite{Batell:2009yf,Batell:2009di} the ratio of the partial widths for 
the Dalitz decays of $\pi^0$, $\eta$ mesons  to $U$-bosons and real photons
can be evaluated as follows:
\begin{eqnarray}
    \frac{\Gamma_{m\rightarrow\gamma U}}{\Gamma_{m\rightarrow\gamma\gamma}} = 2\epsilon^2|F_m(q^2 = M_U^2)|\frac{\lambda^{3/2}(m_m^2, m_\gamma^2, M_U^2)}{\lambda^{3/2}(m_m^2, m_\gamma^2, m_\gamma^2)},
\label{GgU}    
\end{eqnarray}
for $m=\pi^0, \eta$. Here  $\epsilon^2$ is the kinetic mixing parameter and $\lambda$ is the triangle function 
($\lambda(x,y,z)=(x-y-z)^2-4yz$) from the expression of particle 3-momentum. Since $m_\gamma=0$,
one obtains:
\begin{eqnarray}
    \frac{\lambda^{3/2}(m_m^2, 0, M_U^2)}{\lambda^{3/2}(m_m^2, 0, 0)} = \left(  1 - \frac{M_U^2}{m_m^2}\right)^3.
\label{lam}    
\end{eqnarray}
Here, $M_U$ is the $U$-boson mass, $F_m$ are the electromagnetic transition formfactors for $\pi^0$ and $\eta$; they are taken as  
 in our previous studies \cite{Bratkovskaya:2007jk,Bratkovskaya:2013vx} 
 and in the experimental HADES study \cite{Agakishiev:2013fwl} as well:
\begin{eqnarray}
    |F_{\pi^0}(q^2)| = 1+ 0.032 \frac{q^2}{m_{\pi^0}^2}\\
    |F_\eta(q^2)| = \left(1 -  \frac{q^2}{\Lambda^2}\right)^{-1}
\label{Fpi0}    
\end{eqnarray}
\noindent with $\Lambda$ = 0.72\,GeV.

The $U$-boson production by the $\Delta$ Dalitz decay $\Delta \to NU$ has been proposed in Ref. \cite{Batell:2009di}.
For the evaluation of the partial decay widths of a broad $\Delta$  resonance, one has to take into account
the $\Delta$ spectral function $A(M_\Delta)$ as used also in the HADES study \cite{Agakishiev:2013fwl}:
\begin{eqnarray}
&&   \frac{\Gamma_{\Delta\rightarrow NU}}{\Gamma_{\Delta\rightarrow N\gamma}}  \label{DNU} \\
  &&= \epsilon^2\int A(M_\Delta)|F_\Delta(M_U^2)|\frac{\lambda^{3/2}(M_\Delta^2, m_N^2, M_U^2)}{\lambda^{3/2}(M_\Delta^2, m_N^2, m_\gamma^2)} dM_\Delta, \nonumber   
\end{eqnarray}
where  $M_\Delta$ is the mass of the $\Delta$ resonance distributed according to the spectral function $A(M_\Delta)$,
$m_N$ the mass of the remaining nucleon. 
Following Ref. \cite{Agakishiev:2013fwl} we adopted $|F_\Delta(M_U^2)| = 1$ for this study
since an experimental formfactor is unknown.

In the PHSD the spectral function of a $\Delta$ resonance of mass $M_\Delta$
is taken in the relativistic Breit-Wigner form \cite{Bratkovskaya:2013vx}:
\begin{eqnarray}
A_\Delta(M_\Delta) = C_1\cdot {2\over \pi} \ {M_\Delta^2 \Gamma_\Delta^{tot}(M_\Delta)
\over (M_\Delta^2-M_{\Delta 0}^2)^2 + (M_\Delta {\Gamma_\Delta^{tot}(M_\Delta)})^2}.
\label{spfunD}
\end{eqnarray}
with $M_{\Delta 0}$ being the pole mass of the $\Delta$.
The factor $C_1$ is fixed by the normalization condition:
\begin{eqnarray}
\int_{M_{min}}^{M_{lim}} A_\Delta(M_\Delta) dM_\Delta =1,
\label{SFnorma}\end{eqnarray} where $M_{lim}=2$~GeV is chosen as
an upper limit for the numerical integration. The lower limit for
the vacuum spectral function corresponds to the nucleon-pion decay,
$M_{min}=m_\pi+m_N$. In  $NN$ collisions the $\Delta$'s can be populated up to the $M_{max}=\sqrt{s}-m_N$ and hence
the available part of the spectral function depends on the collision energy.

We recall that for the total decay width of the $\Delta$ resonance $\Gamma_\Delta^{tot} (M_\Delta)$ in the PHSD  
we adopt the "Monitz model" \cite{Koch:1983tf} (cf. also Ref. \cite{Wolf:1990ur}):
\begin{eqnarray}
&&\Gamma_\Delta^{tot} (M_\Delta)= \Gamma_R {M_{\Delta 0} \over M_\Delta}
       \cdot \left(q\over q_r\right)^3 \cdot F^2(q), \label{WidthDel}\\
&& q^2={(M_\Delta^2 -(m_N+m_\pi)^2)(M_\Delta^2 -(m_N-m_\pi)^2)
       \over 4 M_\Delta^2}, \nonumber \\
&&  \Gamma_R = 0.11 {\ \rm GeV}, \ \ M_{\Delta 0} = 1.232 {\ \rm GeV}; \nonumber \\
&& F(q) ={\beta_r^2 +q_r^2 \over \beta_r^2 +q^2}, \label{WidthDel1} \\
&&  q_r^2 = 0.051936, \ \ \beta_r^2 = 0.09.\nonumber
\end{eqnarray}
We note that when accounting for the mass-dependent total width of the $\Delta$ resonance  our calculation for
$U$-boson production by the $\Delta$ Dalitz decay differs from the evaluation in Ref. \cite{Agakishiev:2013fwl}
where a constant total width of the $\Delta$ has been used. As discussed in 
Ref. \cite{Bratkovskaya:2013vx} (see Section VI), the shape of the spectral function strongly depends on the
actual form of $\Gamma_\Delta^{tot} (M_\Delta)$.

The branching ratio for the decay of $U$-bosons to $e^+e^-$, entering Eq. (\ref{NUee}), is 
adopted from Ref. \cite{Batell:2009yf} and used also in Ref. \cite{Agakishiev:2013fwl}:
\begin{eqnarray}
   Br^{U\to ee} &&= \frac{\Gamma_{U \rightarrow e^+e^-}}{\Gamma_{tot}^U} \label{Bree}\\
&&     = \frac{1}{1 + \sqrt{1 - \frac{4m_\mu^2}{M_U^2}} \left( 1 + \frac{2m_\mu^2}{M_U}\right) \left(1 + R(M_U)\right)}.
\nonumber    
\end{eqnarray}
Here $m_\mu$ is the muon mass. The total decay width of a $U$-boson is the sum of the partial decay widths 
to hadrons, $e^+e^-$ and $\mu^+\mu^-$ pairs:
$\Gamma_{tot}^U= \Gamma_{U\to hadr} + \Gamma_{U\to e^+e^-} + \Gamma_{U\to\mu^+\mu^-}$.
The expression (\ref{Bree}) has been evaluated using that $\Gamma_{U\to\mu^+\mu^-} = \Gamma_{U\to e^+e^-}$ 
due to lepton universality for $M_U\gg 2m_\mu$. The hadronic decay widths of $U$-bosons 
is chosen such that $\Gamma_{U\to hadr} = R(\sqrt{s}=M_U)\Gamma_{U\to\mu^+\mu^-}$, where 
the factor $R(\sqrt{s}) = \sigma_{e^+e^-\rightarrow hadrons}$/$\sigma_{e^+e^-\rightarrow \mu^+\mu^-}$ 
is taken from Ref. \cite{Beringer:1900zz}.

\section{Results for the dilepton spectra from U-boson decays and constraints on $\epsilon^2 (M_U)$}

In this Section we present the numerical results within the PHSD approach for the dilepton spectra 
including the contribution from the dilepton decay of $U$-bosons.

Since the kinetic mixing parameter $\epsilon^2$  is unknown as well as the mass of the $U$-boson, we invent the
following procedure to obtain the constraints on $\epsilon^2 (M_U)$:
for each bin in dilepton mass $dM$, which is taken to be 10 MeV in our simulations,
we calculate the integrated yield of dileptons from $U$-bosons of masses $[M_U,M_U+dM]$
according to Eq. (\ref{NUee}) and divide by the bin size $dM$. The resulting dilepton yield
per bin $dM$ we denote as $dN^{sumU}/dM$, which is the sum of all contributions (kinematically possible in the
mass bin) from the dilepton decay of $U$-bosons produced by the Dalitz decays of $\pi^0$, $\eta$ 
and the $\Delta$ resonance.
Assuming that $\epsilon^2$ is a constant in $dM$ we can write that 
$dN^{sumU}/dM=\epsilon^2 dN^{sumU}_{\epsilon=1}/dM$
where the notation $dN^{sumU}_{\epsilon=1}/dM$ is the dilepton yield
calculated without $\epsilon^2$ or formally with $\epsilon=1$.

Thus, the total sum of all possible sources of dileptons, from the SM channels and 
from $U$-boson decays, can be written as 
\begin{eqnarray}
  \frac{dN}{dM}^{total} &=& \frac{dN}{dM}^{sum SM} + \frac{dN}{dM}^{sum U} \nonumber\\
   &=& \frac{dN}{dM}^{sum SM} + \epsilon^2\frac{dN_{\epsilon=1}^{sum U}}{dM}.
\label{dNdMepsil}
\end{eqnarray}
Now we can obtain constraints on $\epsilon^2 (M_U)$ by requesting that the total sum $dN^{total}/dM$
cannot surplus the sum of SM channels by more than a fraction $C_U$ in each bin $dM$,
i.e.\ $C_U$ controls the additionally "allowed" di-electron yield resulting from dark photons on top of the total SM yield 
(e.g. $C_U=0.1$ indicating that the dark photons add 10\% extra yield to the SM yield, $C_U=0.2$ meaning 20\% extra, etc.).
We then express this as
\begin{eqnarray} 
  \frac{dN}{dM}^{total} = (1+C_U) \frac{dN}{dM}^{sum SM}.
\label{dNdMepsil1}
\end{eqnarray}
Combining Eqs. (\ref{dNdMepsil1}) and  (\ref{dNdMepsil}), one obtains that 
the kinetic mixing parameter $\epsilon^2$ for $M_U$ can be evaluated as 
\begin{eqnarray}
    \epsilon^2 (M_U) = C_U \cdot  \left. { \left(\frac{dN}{dM}^{sumSM} \right)} \right/
    {\left(\frac{dN_{\epsilon=1}^{sum U}}{dM} \right)}.
\label{epsM}
\end{eqnarray}
Eq. (\ref{epsM}) allows to compute $\epsilon^2$ for each bin $[M_U,M_U+dM]$ and presents the properly weighted 
dilepton yield from dark photons relative to the SM contributions. Moreover, now we can apply the experimental 
acceptance for $e^+e^-$ pairs from $U$-boson decays in the same way as for the SM channels and compare 
our results to the experimental data from the HADES Collaboration. The latter will allow to explore the possible range
of the factor $C_U$ that controls the additional yield from dark photons to the SM contributions.
Since the dark photons have not been observed in any dilepton experiments, one can require that 
this enhancement should be still in the acceptable agreement with experimental data, i.e. 
within the experimental error bars (under condition that the SM yield agrees well with experimental data).

\begin{figure*}[!]
\centerline{
\includegraphics[width=8.1 cm]{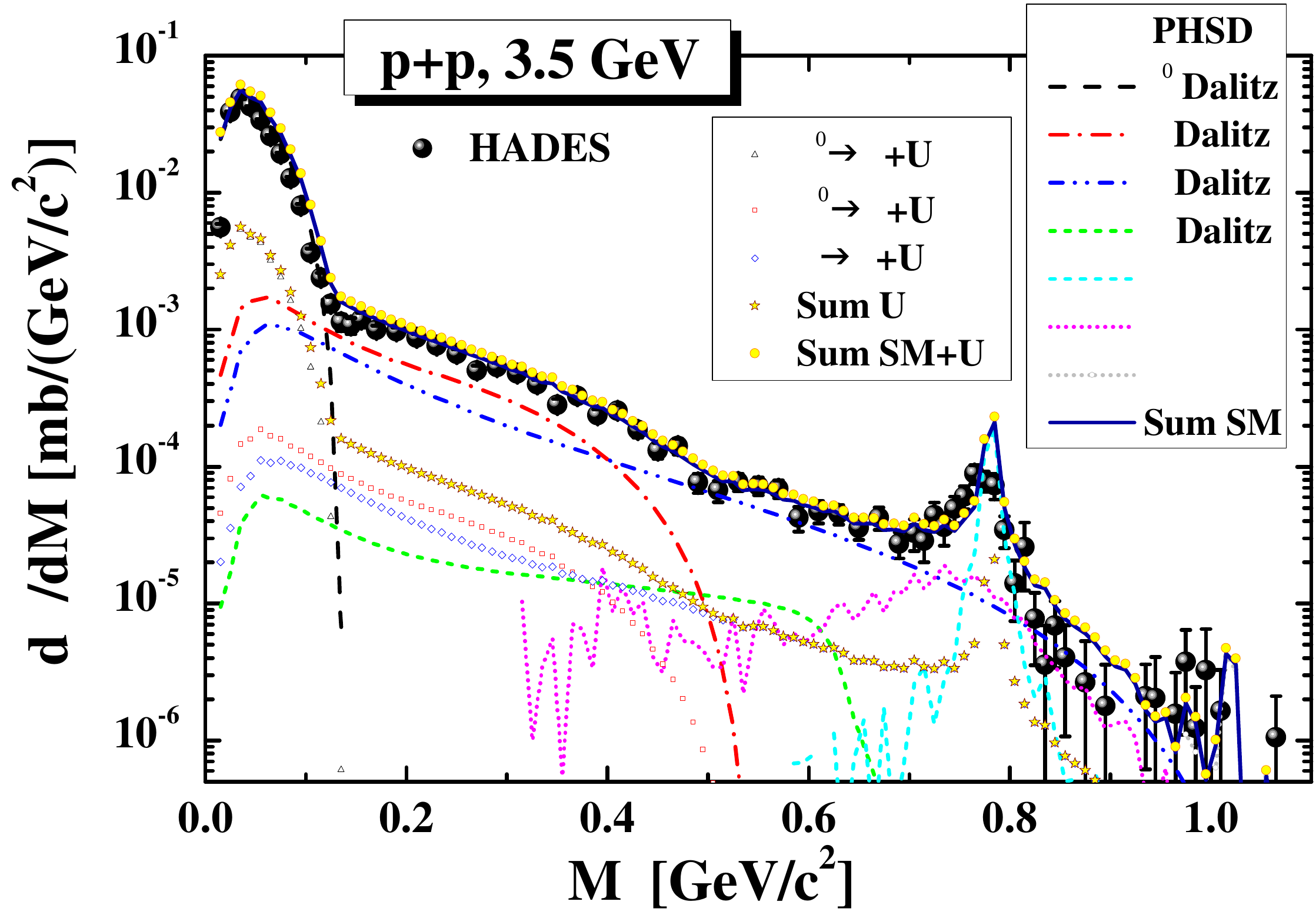}\hspace*{5mm}
\includegraphics[width=8.1 cm]{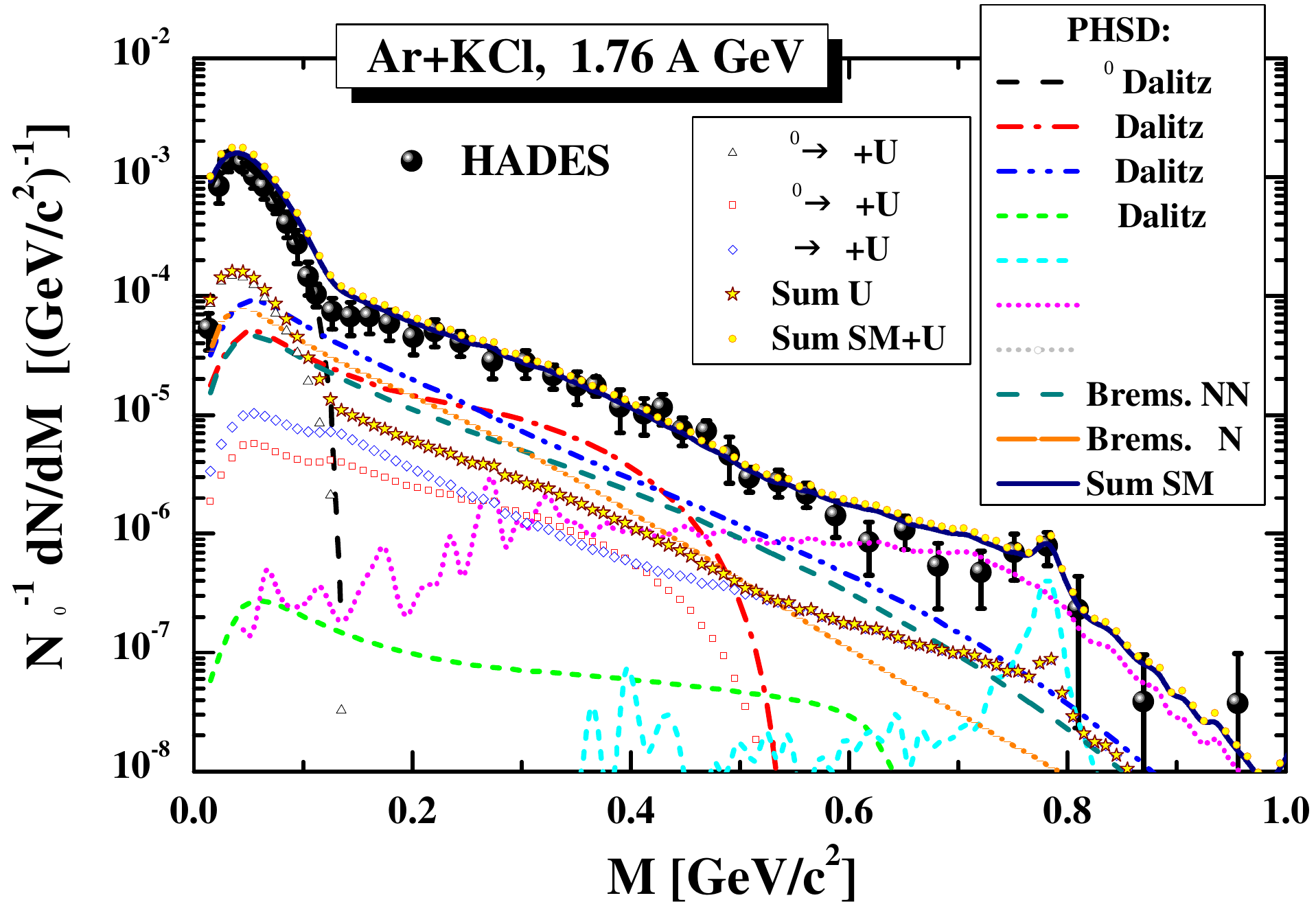}}
\centerline{
\includegraphics[width=8.1 cm]{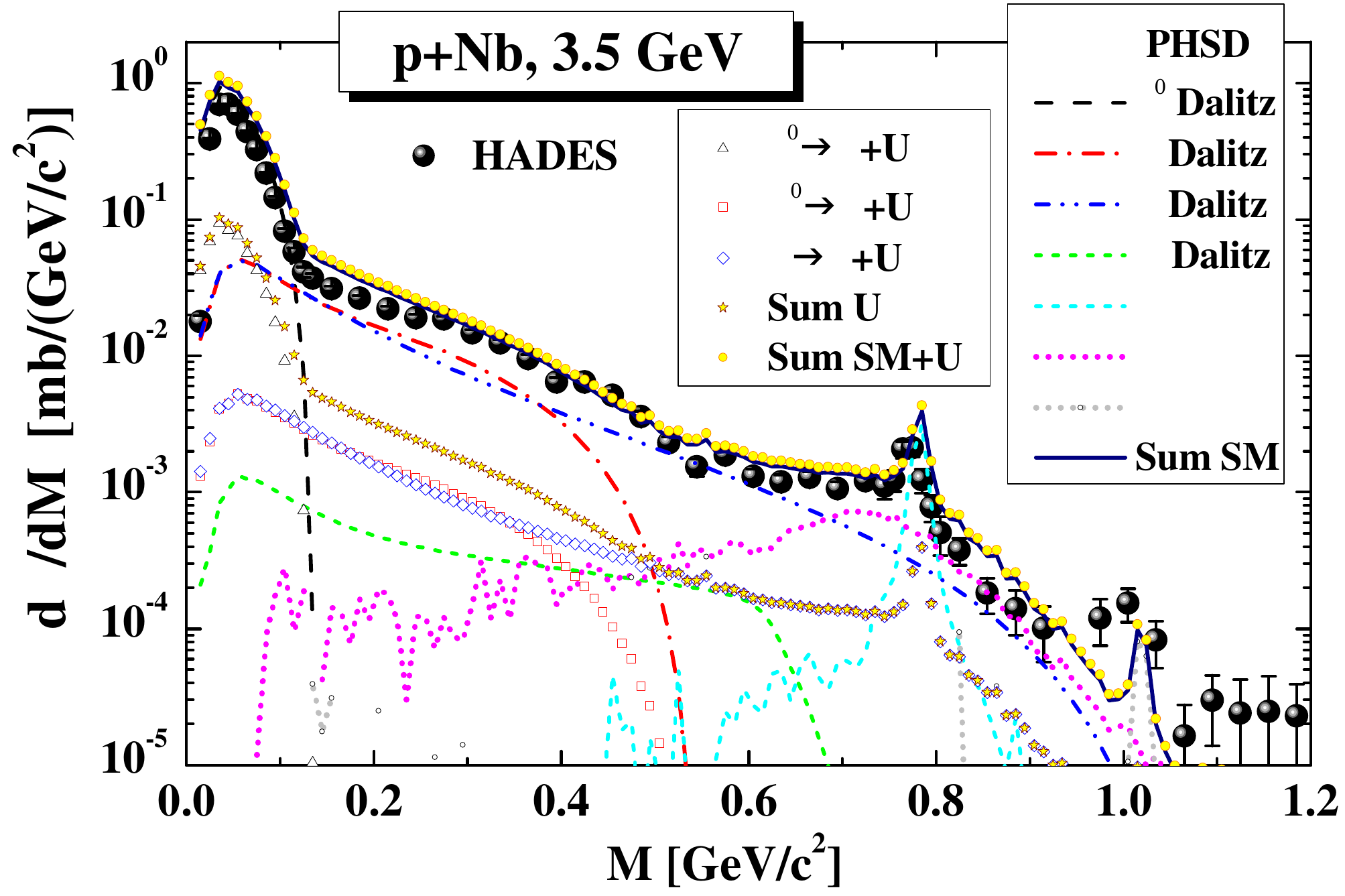}\hspace*{5mm}
\includegraphics[width=8.1 cm]{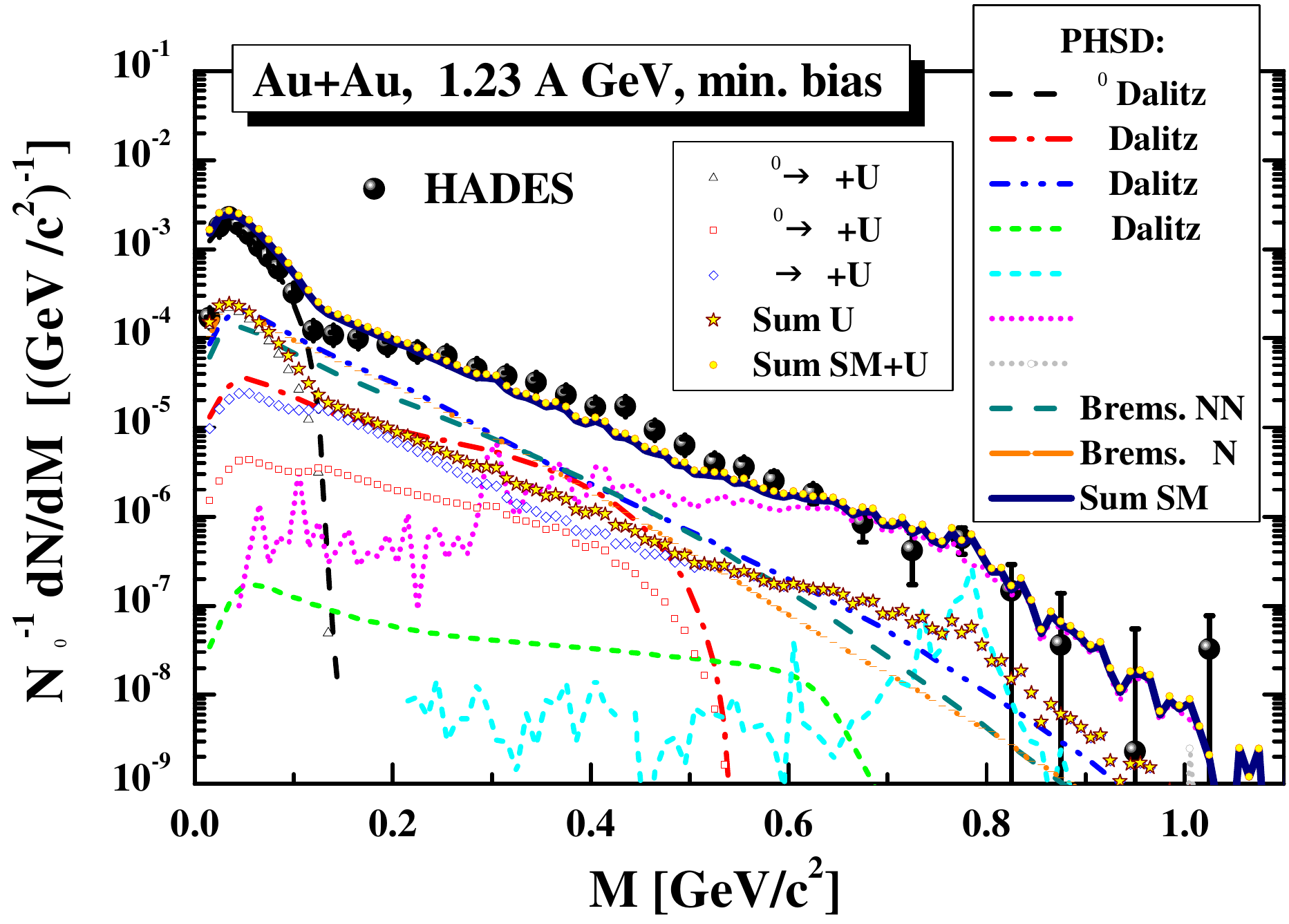}}
\caption{The PHSD results for the differential cross section $d\sigma/dM$ for $e^+e^-$ production in $p+p$ (upper, left) and $p+Nb$ reactions (upper, right) at 3.5 GeV beam energy and 
for the mass differential dilepton spectra $dN/dM$ - normalized to the $\pi^0$ multiplicity - for
$Ar+KCl$ collisions at 1.76\,A GeV (lower, left) and for $Au+Au$ collisions at 1.23 A GeV 
(lower, right) in comparison to the experimental measurements by the HADES Collaboration.
The solid dots present the HADES data for $p+p$ \cite{HADES:2011ab}, for $p+Nb$ \cite{Weber:2011zze,Agakishiev:2012vj},
for $Ar+KCl$ \cite{Agakishiev:2011vf} and for $Au+Au$ \cite{Adamczewski-Musch:2019byl}, respectively. 
The individual colored lines display the contributions from the various SM channels of dilepton production
in the PHSD calculations (cf. color coding in the legend). 
The contributions from $U \rightarrow e^+e^-$ (with 10\% allowed surplus of the total SM yield) 
produced by Dalitz decays of $\pi^0$ are shown as black triangles,
of $\eta$ as red squared, of $\Delta$-resonance as blue rhombus, their sum - as brown stars and the sum of dileptons from
all  SM channels and U-decays - as yellow dots.
The theoretical calculations are passed through the corresponding HADES
acceptance filter and mass/momentum resolution. }
\label{M_Hades}
\end{figure*}

In Fig. \ref{M_Hades} we present the compilation of the PHSD results for 
the differential cross section $d\sigma/dM$ for $e^+e^-$ production in $p+p$ (upper, left) 
and $p+Nb$ reactions (upper, right) at 3.5 GeV beam energy and 
for the mass differential dilepton spectra $dN/dM$ -- normalized to the $\pi^0$ multiplicity -- for
$Ar+KCl$ collisions at 1.76\,A GeV (lower, left) and for $Au+Au$ collisions at 1.23 A GeV 
(lower, right) in comparison to the experimental measurements by the HADES Collaboration.
The solid dots present the HADES data for $p+p$ \cite{HADES:2011ab}, for $p+Nb$ \cite{Weber:2011zze,Agakishiev:2012vj}
for $Ar+KCl$ \cite{Agakishiev:2011vf} and for $Au+Au$ \cite{Adamczewski-Musch:2019byl}. 
We note that the theoretical calculations are passed through the corresponding HADES
acceptance filter and mass/momentum resolution.

The contributions from the various SM channels of dilepton production in the PHSD calculations 
are presented as individual colored lines described in the legend:
the black dashed lines show the contribution from $\pi^0$ Dalitz decays, the red dot-dashed lines --
 from $\eta$ Dalitz decays, the blue dot-dot-dashed lines --  from $\Delta$ Dalitz decays, 
 the green short dashed lines -- from $\omega$ Dalitz decays, 
 the light blue short dashed  lines --  from direct decay $\omega\to e^+e^-$ ,
 the magenta dotted lines --  from direct decay $\rho\to e^+e^-$, 
 the grey dotted lines with open dots --  from direct decay $\phi\to e^+e^-$, 
 the navy solid lines correspond to the sum of all SM contributions ("Sum SM" in legend).
We note that for low beam energies ($E_{kin}<2$ A GeV) we account also for the Bremsstrahlung contributions
(as described in Section II.B):  
the dark cyan dashed lines show the $NN=pn+pp$ Bremsstrahlung and orange wave lines indicate 
the $\pi N$ Bremsstrahlung. However, we do not include the Bremmsstrahlung contributions for larger energies
since the OBE calculations used for low energies can not be easily extrapolated to the 
high energy regime. As has been shown in Ref. \cite{Bratkovskaya:2013vx}, such an extrapolation
leads to a slight overestimation of the experimental data.
We stress that we include the collisional broadening scenario for the vector meson ($\rho, \omega, \phi $) 
spectral functions which leads to a smearing of the peaks in the case of heavy-ion collisions and 
to an approximate exponential shape of the dilepton yields for heavy systems such as Au+Au collisions 
compared to the clear peak structure for $p+p$ collisions. 
As seen from   Fig. \ref{M_Hades}, the PHSD results for the SM sources are in a good agreement with the
HADES results for all four systems as well as with the previous PHSD results in Ref. \cite{Bratkovskaya:2013vx}.
 
The individual contributions from $U$-boson decay to dileptons are shown in Fig. \ref{M_Hades}
as symbols: the contributions from $U \to e^+e^-$ produced by Dalitz decays of $\pi^0$ are shown by black triangles,
from $\eta$ -- as red squared, from $\Delta$ -- as blue rhombus. The sum of all DM channels (allowed
kinematically in a given $M=M_U$ bin) is shown by brown stars (indicated as "Sum U" in the legend) 
and the sum of dileptons from all SM channels and U-decays is displayed  as yellow dots ("Sum SM+U" in legend).
The calculation of dileptons from $U$-bosons have been done with the factor $C_U=0.1$, i.e. by allowing 
10\% surplus of the theoretical results from the total sum of all SM channels.

\begin{figure}[!]
\centerline{
\includegraphics[width=8.cm]{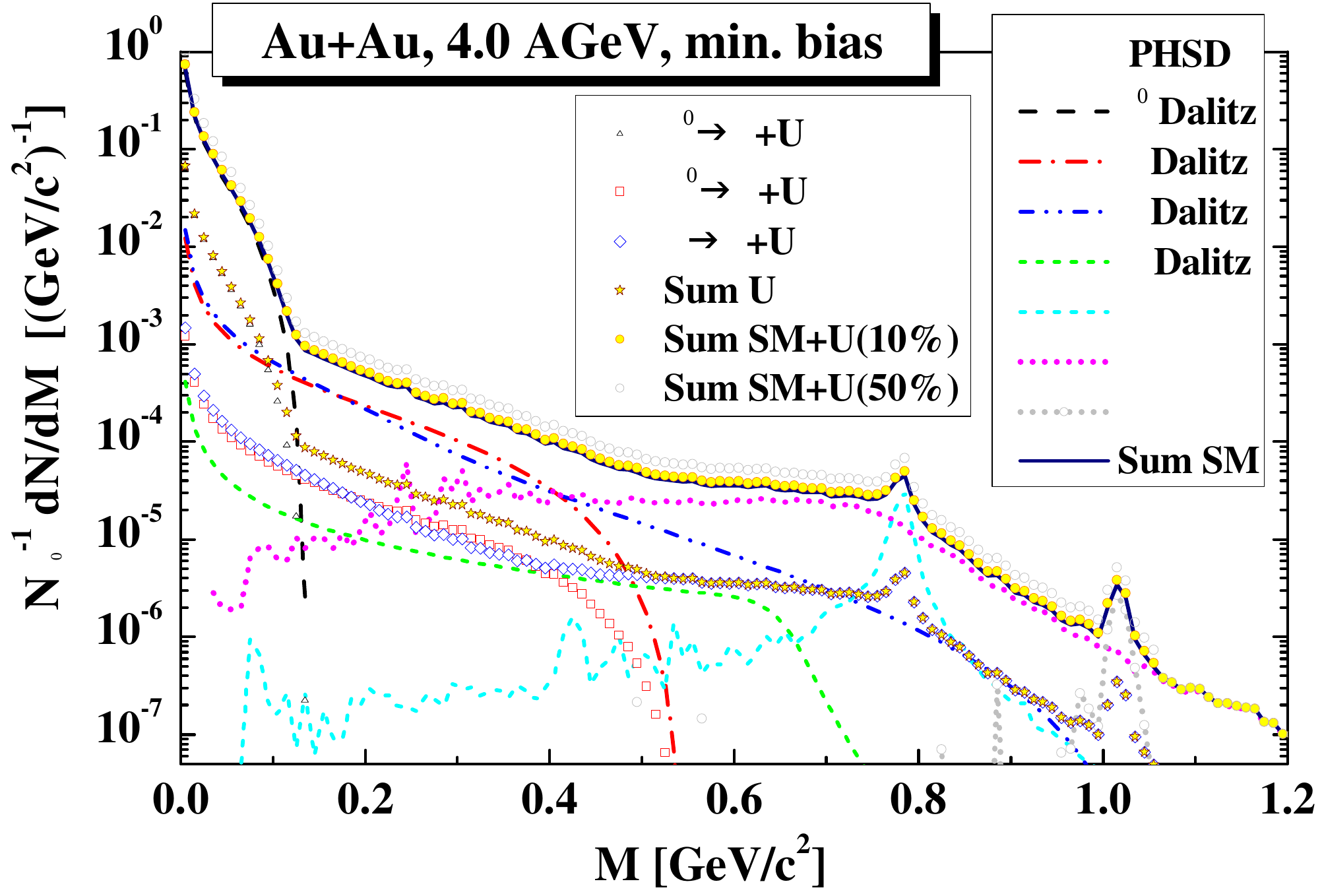}}
\caption{ The PHSD predictions for the mass differential dilepton spectra $dN/dM$ -
normalized to the $\pi^0$ multiplicity - for minimum bias $Au+Au$ collisions at 4.0 A GeV. 
The color coding of the individual lines is the same as in Fig. \ref{M_Hades}.
The contributions from $U \rightarrow e^+e^-$  are shown with 10\% allowed surplus of total SM yield. 
Additionally, the grey open dots show the sum of all SM and $U$-boson contributions with 50\% 
allowed surplus of the total SM yield. }
\label{AuAu40}
\end{figure}

In Fig. \ref{AuAu40} we show the PHSD result for the mass differential dilepton spectra $dN/dM$ -
normalized to the $\pi^0$ multiplicity - for the SM and DM channels 
for minimum bias Au+Au collisions at 4.0 A GeV which is a prediction for future FAIR and NICA experiments. 
The description of the individual lines are the same as in Fig. \ref{M_Hades}.
Here we do not apply any acceptance cuts. Also we consider for the dark photons $C_U=0.1$, i.e. 
10\% allowed surplus of the total SM yield. Additionally, we show the sum of 
all $U$-boson contributions with 50\% allowed surplus of the  SM yield by the grey open dots.

As follows from Figs. \ref{M_Hades} and  \ref{AuAu40} the total contributions from DM sources
follow the shape of the corresponding SM contributions. This follows from the constraints on $\epsilon^2(M_U)$
given by Eq. (\ref{epsM}) which "scale" the DM yield by the same factor $C_U$ in each $M_U$ bin.
One can see that with increasing beam energies the contribution of vector mesons becomes more and more important
for $M > 0.6$ GeV. As advocated in Refs. \cite{Batell:2009yf,Batell:2009di} the $U$-bosons can be also
produced by vector meson conversion due to the $U(1)-U(1)^\prime$ mixing; these channels are not accounted for
in the present study, however,  are a subject of future investigation.

\begin{figure}[!]
\centerline{
\includegraphics[width=8. cm]{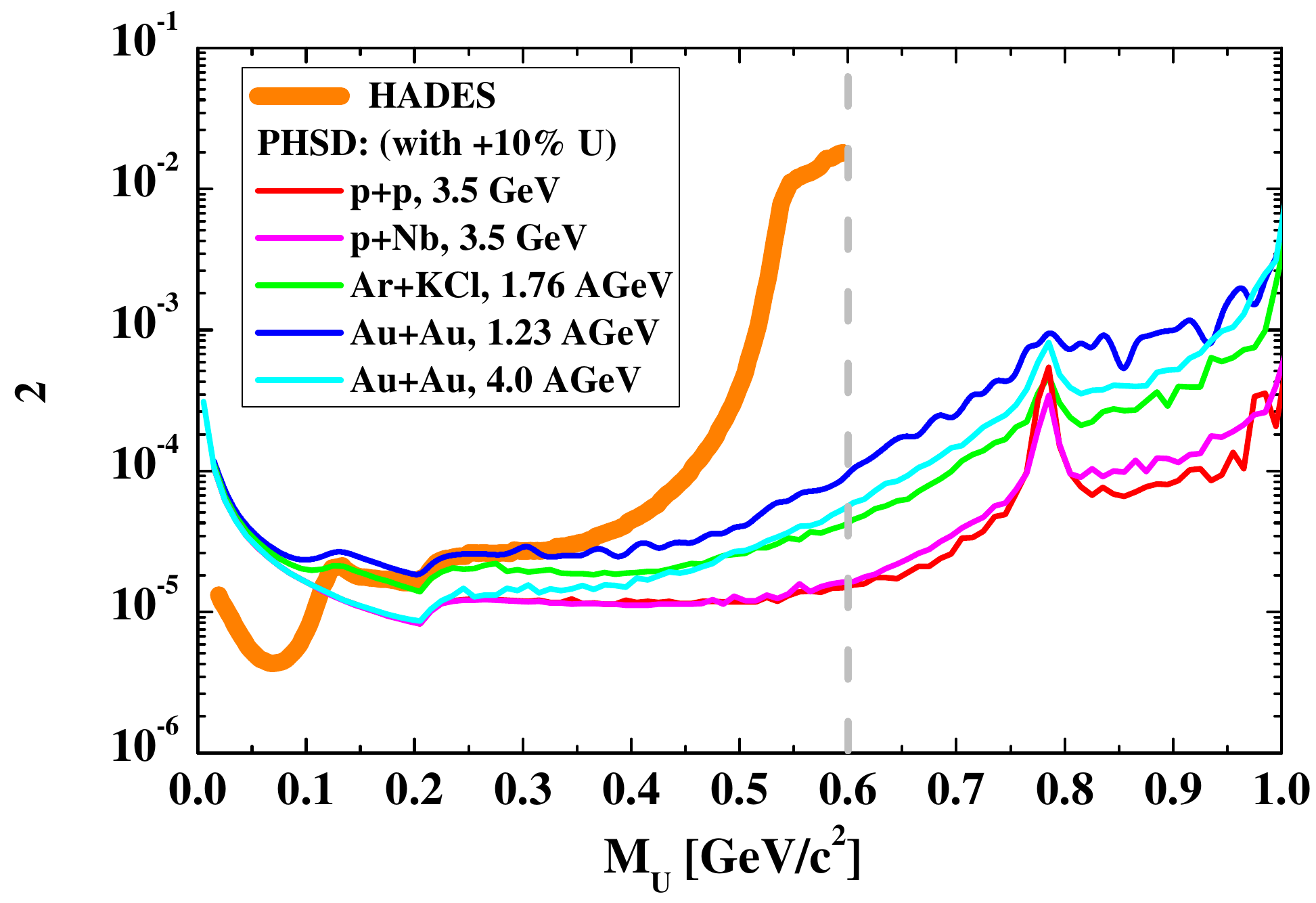}}
\caption{
The kinetic mixing parameter $\epsilon^2$ extracted from the PHSD dilepton spectra for $p+p$ at 3.5 GeV (red line), 
$p+Nb$ at 3.5 GeV (magenta line),  $Ar+KCl$ at 1.76 A GeV (green line), $Au+Au$ at 1.23 A GeV (blue line),
$Au+Au$ at 4.0 A GeV in comparison with the combined HADES results (orange line) \cite{Agakishiev:2013fwl}.
The PHSD results are shown with 10\% allowed surplus of the $U$-boson contributions over the total SM yield, i.e.\ $C_U  = 0.1$.
}
\label{epsil2}
\end{figure}

Finally, in Fig. \ref{epsil2} we show the results for the kinetic mixing parameter $\epsilon^2$ versus $M_U$
extracted from the PHSD dilepton spectra for $p+p$ at 3.5 GeV (red line), 
$p+Nb$ at 3.5 GeV (magenta line),  $Ar+KCl$ at 1.76 A GeV (green line), $Au+Au$ at 1.23 A GeV (blue line),
$Au+Au$ at 4.0 A GeV in comparison with the combined HADES results (orange line) from Ref. \cite{Agakishiev:2013fwl}.
The PHSD results are shown with 10\% allowed surplus of the $U$-boson contributions over the total SM yield.
The grey dashed vertical line limits the region of applicability of our estimates of $\epsilon^2$, i.e.
$M_U\le 0.6$ GeV, since at larger dilepton masses the contributions from vector mesons become important.
However, we still show our calculations for larger $M_U$ to demonstrate that our theoretical method can provide 
useful constraints on $\epsilon^2$ for any $M_U$. 
Note also that the shape of our theoretically obtained $\epsilon^2(M_U)$ curve is not affected by the experimental 
detector acceptance since the latter acts in the same manner on the SM and DM contributions at a given $M=M_U$.
One can see that the extracted $\epsilon^2$ for small $(M_U) < m_{\pi^0}$ depends only slightly on the size of the collision system
and energy since, the $\pi^0$ Dalitz decay is the dominant channel.  With increasing $M_U$ more channels
are open and the result depends on the fraction of the $U$-boson production channels relative to all other dilepton channels.
 
As follows from  Fig. \ref{epsil2}, our $\epsilon^2(M_U)$ results calculated with $C_U = 0.1$ are close to the HADES
experimental result  \cite{Agakishiev:2013fwl} for $0.15 < M_U < 0.4$ GeV. 
We point out that this HADES finding is above the upper limit provided in the
compilation of  world wide experimental results \cite{Agrawal:2021dbo}: 
the measurements by the NA48/2 \cite{Batley:2015lha}, BaBar \cite{Lees:2014xha,Lees:2017lec} 
and A1  \cite{Merkel:2014avp} experiments set the present world limit at about $10^{-6}$ in this mass range.

\begin{figure}[!]
\centerline{
\includegraphics[width=8.cm]{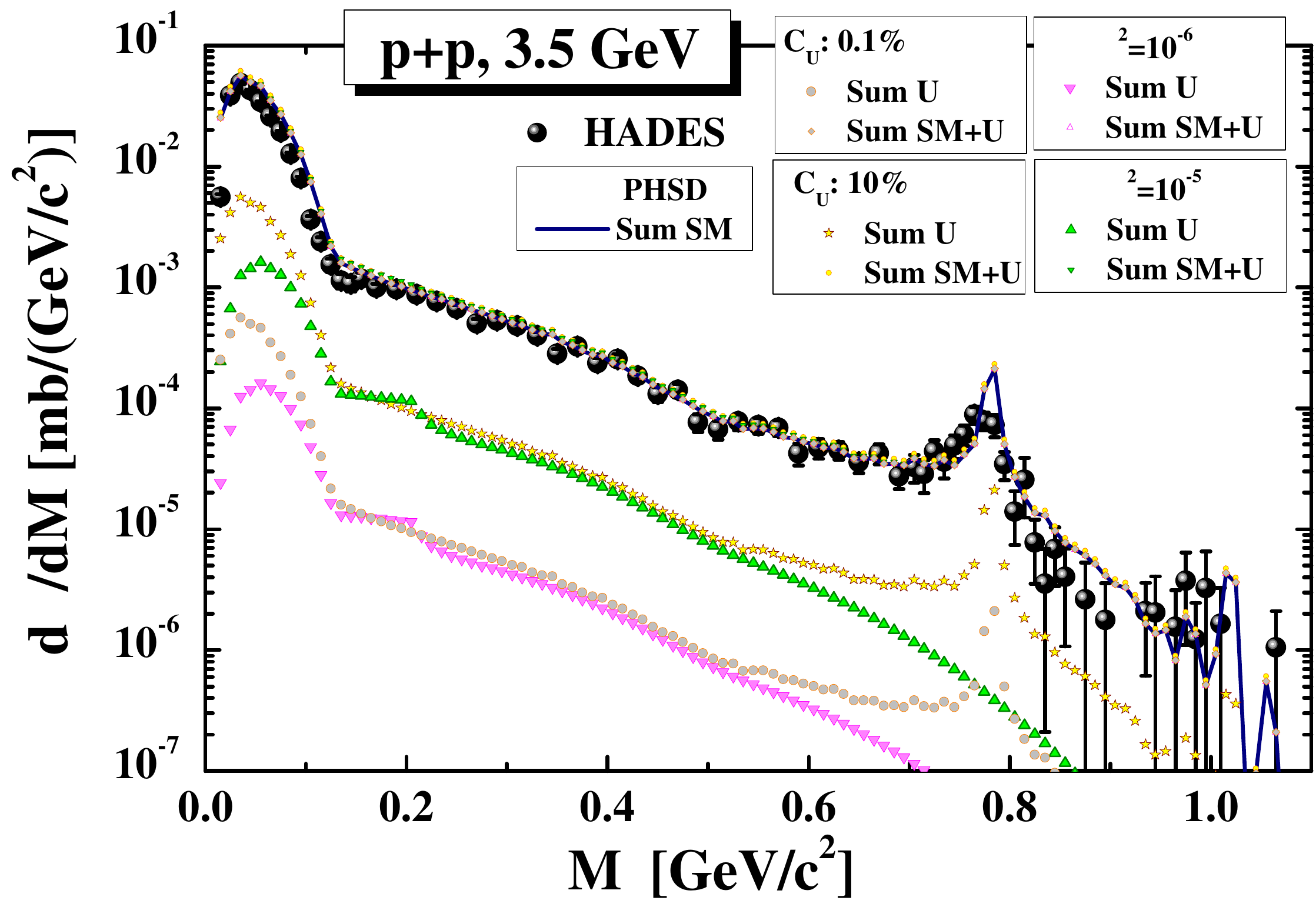}}
\caption{The PHSD results for the differential cross section $d\sigma/dM$ 
for $e^+e^-$ production in $p+p$ reactions at 3.5 GeV beam energy calculated for different 
$\epsilon^2$ scenarios  in comparison to the experimental measurements by the HADES Collaboration \cite{HADES:2011ab}: i) $C_U$ corresponding to  0.1\% and 10\% (as in Fig. \ref{M_Hades}) allowed surplus of total SM yield; ii) constant 
$\epsilon^2 =10^{-5}$ and $\epsilon^2 =10^{-6}$ (cf. color coding in the legend).  }
\label{pp35ConstE}
\end{figure}

\begin{figure}[!]
\centerline{
\includegraphics[width=8. cm]{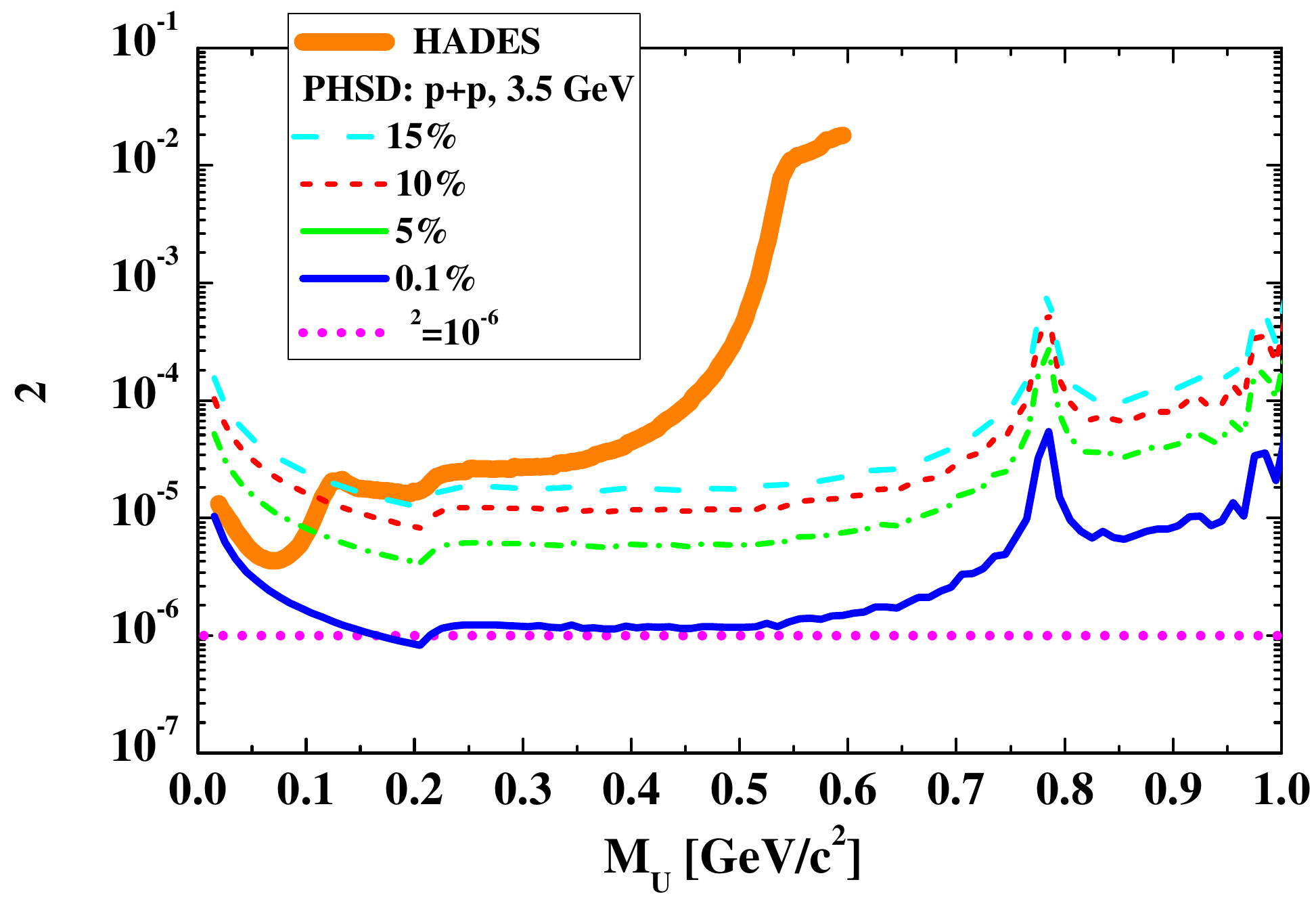}}
\caption{
The kinetic mixing parameter $\epsilon^2$ extracted from the PHSD dilepton spectra for $p+p$ at 3.5 GeV  
calculated for different $\epsilon^2$ scenarios in comparison with the combined HADES results (orange line) \cite{Agakishiev:2013fwl} (as in Fig. \ref{epsil2}).
The PHSD results are shown with 0.1, 5, 10, 15\% allowed surplus of the $U$-boson contributions over 
the total SM yield (cf. color coding in the legend). The dotted magenta line shows the constant 
$\epsilon^2 =10^{-6}$. }
\label{epsil2C}
\end{figure}

In order to explore the theoretical uncertainties in setting the upper limit we present in Fig. \ref{pp35ConstE}
(similar to Fig. \ref{M_Hades}) - the PHSD  results for the differential cross section $d\sigma/dM$ 
for $e^+e^-$ production in $p+p$ reactions at 3.5 GeV beam energy calculated for different 
$\epsilon^2$ scenarios  in comparison to the experimental measurements by the HADES Collaboration \cite{HADES:2011ab} : i) $C_U$ corresponding to  0.1\% and 10\% (10\% as in Fig. \ref{M_Hades}) allowed surplus of total SM yield; ii) constant $\epsilon^2 =10^{-5}$ and $\epsilon^2 =10^{-6}$ (cf. color coding in the legend). 
We note that the shape of the $d\sigma/dM$ dileptons from $U$-bosons decays 
shown in  Fig. \ref{pp35ConstE} for $\epsilon^2 =10^{-6}$ is different from the shape of the same contributions
presented in Fig. \ref{M_Hades} for $C_U=0.1$ (i.e. 10\% surplus over the total SM channels). 
One can clearly see the 'step' behaviour at the mass $\sim 0.21$ GeV which reflects the mass dependence of 
the branching ratio for the decay of $U$-bosons to $e^+e^-$ (Eq. (\ref{Bree}))
and is related to the opening of the $\mu^+\mu^-$ channel (cf. Fig. 1 of Ref. \cite{Agakishiev:2013fwl}).
We find that the $dN/dM$ spectra calculated with constant $\epsilon^2 =10^{-6}$ are coincident 
with those calculated with $C_U=0.001$ which is 0.1\% surplus in the $0.15 < M < 0.6$ GeV mass range.
On the other hand, the 10\% surplus - shown in Figs. \ref{M_Hades} and \ref{pp35ConstE} and discussed above -
corresponds to a constant $\epsilon^2 =10^{-5}$ in the $0.15 < M < 0.6$ GeV mass range.
We mention that both scenarios for the extraction of the upper limit of the mixing parameters are based on the
comparison of theoretical results - assuming that the theoretical model reproduces the SM contributions "ideally"
which is not exactly the case. The underlying uncertainties come from the calculation of the Dalitz decays
of mesons and $\Delta$ resonances due to the lack of knowledge of the corresponding formfactors 
as well as a relative contribution of individual channels (cf. discussions in Section II).

For the illustration of the discussion above we show in Fig. \ref{epsil2C} the kinetic mixing parameter $\epsilon^2$ 
extracted from the PHSD dilepton spectra for $p+p$ at 3.5 GeV  calculated for different $\epsilon^2$ scenarios in comparison with the combined HADES results (orange line) \cite{Agakishiev:2013fwl} (as in Fig. \ref{epsil2}).
The PHSD results are shown with 0.1, 5, 10, 15\% allowed surplus of the $U$-boson contributions over 
the total SM yield (cf. color coding in the legend). The dotted magenta line shows the constant 
$\epsilon^2 =10^{-6}$ which approximately corresponds to the present knowledge on the upper limit 
based on the compilation of the world  wide experiments \cite{Agrawal:2021dbo}.

We have to stress that the leverages applied to the extraction of an upper limit for $\epsilon^2(M_U)$ in our calculations or in
the HADES experiment are rather different:  while in our theoretical approach, the excess caused by a hypothetical $U$-boson is shown as
a function of the (yet unkown) $U$-boson mass (for a given fixed probability taken here as 10\%),  the experimental approach is to search for
an unexpected peak structure of given width (defined by the detector resolution) on top of a smooth continuum.
This also means that in the experiment the sensitivity to find such a peak is basically limited by the statistical fluctuations of the
measured total dilepton yield.  Except for the $\pi^0$ Dalitz region the measured yield is dominated by the irreducible combinatorial 
background,
i.e.\ the backgorund of uncorrelated lepton pairs.  Ultimately, the pair statistics determines the uncertainties at higher invariant masses and
the only way to reduce the impact of these fluctuations is by either increasing the detection efficiency, extending the experimental run time and/or
increase the event rates, i.e. run with larger beam intensities, this is planned by the CBM experiment at FAIR \cite{Ablyazimov:2017guv}. 
\section{Summary}

In this study we presented the first microscopic transport calculations, based on the PHSD approach, 
for the dilepton yield from the decay of hypothetical dark photons (or $U$-bosons), $U\to e^+e^-$,
from $p+p$, $p+A$ and heavy-ion collisions at SIS18 energies. 
For that we incorporated in the PHSD the production of $U$-bosons by the Dalitz decay of pions 
$\pi^0\to \gamma U$, $\eta$-mesons $\eta \to \gamma U$ and Delta resonances $\Delta \to N U$ decays
based on the theoretical model by Batell, Pospelov and Ritz from in Refs. \cite{Batell:2009yf,Batell:2009di} 
which describes the interaction of DM and SM particles by the $U(1)-U(1)^\prime$ mixing. 
The strength of these interactions is defined by the kinetic mixing parameter $\epsilon^2$ 
which is a parameter in the model. Moreover, the mass of the $U$-boson $M_U$ is also unknown,
i.e. $\epsilon^2=\epsilon^2(M_U)$.

We introduce a procedure to define theoretical constraints on the upper limit of $\epsilon^2(M_U)$:\\
i) Since the dark photons are not observed in dilepton experiments so far (e.g. they were not found
in the HADES dark photon search \cite{Agakishiev:2013fwl}), we can require that their contribution
can not exceed some limit which would make them visible in experimental data. \\
ii) While the dilepton spectra from the SM channels are in a good agreement with the HADES experimental data
for $p+p$, $p+Nb$ at 3.5 GeV,  $Ar+KCl$ at 1.76 A GeV and $Au+Au$ at 1.23 A GeV, we can impose a constraint that
the extra contribution from the $U$-boson decay in each bin of invariant mass $M=M_U$ 
can not exceed the SM yield by more than some small factor $C_U$ which we can vary in our calculations.\\
iii) By setting $C_U$ to e.g. 0.1 (which implies that we "allow" 10\% of extra dark photon yield at each $M_U$),
we obtained constraints on $\epsilon^2(M_U)$ which can be calculated according to Eq. (\ref{epsM})
as a ratio of the total SM yield over the total dark photon yield when setting formally $\epsilon=1$.

This procedure allowed to estimate  the $\epsilon^2(M_U)$ mixing parameter 
in the mass interval $M_U<0.6$ GeV  based on pure theoretical results for dilepton spectra from SM sources and 
dark photon decay with any 'requested' accuracy.
We have demonstrated how the result will change by reducing the surplus
of DM contribution from 0.1\% to 50\% over the SM spectra.

We found that the upper limit of $\epsilon^2(M_U)$ extracted for 10\% surplus of DM contributions 
over SM sources is consistent with the experimental results of the HADES experiment \cite{Agakishiev:2013fwl}, for $0.15 < M_U < 0.4$ GeV, i.e. the calculations are on a level of the experimental upper 
errorbars.  However, the HADES limit on $\epsilon^2(M_U)$ has been superseded by 
the NA48/2, BaBar and A1 experiments: the modern  world wide experimental 
compilation  \cite{Agrawal:2021dbo} shows that  $\epsilon^2 \sim 10^{-6}$ in this mass range.
In fact, the meaning of the quoted upper limit is that  $\epsilon^2 \le 10^{-6}$ 
at a confidence level of 90\%. 
This would lead to only 0.1\% allowed surplus of the DM over SM contributions.
Indeed, such an accuracy is very difficult to achieve in theoretical calculations
due to the model uncertainties as well as statistical fluctuations during the simulations.
Still, we believe that our model allows to run very realistic simulations, in particular of dark photon production in the environment of a heavy-ion collision, useful for optimizing the analysis of existing data and for planning future, more sensitive measurements.

We note that in the present study we focussed only on light dark photons  with  $M_U<0.6$ GeV since the
considered Dalitz decay channels for $U$-boson production are dominant in this mass range, especially for the low energy reactions as measured by the HADES experiment. However, in future
our study can be extended to  larger $M_U$ masses which would require to account for the production of $U$-bosons by other channels such as direct decays of vector mesons and Drell-Yan type of processes which are
dominant for  dilepton production at  intermediate and high dilepton mass regions.

Furthermore, we predicted limits for the dark photon contribution to the dilepton yield for $Au+Au$ collisions
at 4.0 A GeV, which is of interest
for the upcoming FAIR \cite{Ablyazimov:2017guv} and NICA \cite{Kekelidze:2016hhw} experiments.
Since the experimental procedure of searching for $U$-boson dilepton decays is related
to the observation of a sharp peak on top of a smooth continuum, it is also constrained by statistical fluctuations of the combinatorial background. 
In this respect, it may be more promising to search for  peak structures in dilepton experiments optimized for a maximal signal/background ratio,
i.e.\ by using  $p+A$ reactions or collisions of light nuclei  to 
reduce the total $\pi^0$ multiplicity per event.

\vspace*{1mm}
\section*{Acknowledgements}
The authors acknowledge inspiring discussions with  Marcus Bleicher and Wolfgang Cassing. 
Also we thank the COST Action THOR, CA15213 and CRC-TR 211 'Strong-interaction matter under extreme conditions'- project Nr.315477589 - TRR 211 as well as the European Union’s Horizon 2020 research and innovation program under grant agreement No 824093 (STRONG-2020). 
The computational resources have been provided by the LOEWE-Center for Scientific Computing.

\bibliography{biblio_dark_photons}  

\end{document}